\documentclass{article}

\usepackage{arxiv}
\usepackage[utf8]{inputenc} 
\usepackage[T1]{fontenc}    
\usepackage{hyperref}       
\usepackage{url}            
\usepackage{booktabs}       
\usepackage{amsfonts}       
\usepackage{nicefrac}       
\usepackage{microtype}      
\usepackage{lipsum}
\usepackage{graphicx, multirow}

\title{Mutual Hyperlinking Among \\ Misinformation Peddlers}

\author{
  Vibhor Sehgal\\
  School of Information\\
  University of California, Berkeley \\
  Berkeley, CA \\
  \texttt{sehgalvibhor@berkeley.edu} \\
   \And
  Ankit Peshin \\
  Avast, Inc.\\
  Emeryville, CA \\
  \texttt{ankit.peshin@avast.com} \\
   \And
  Sadia Afroz \\
  Avast, Inc.\\
  Emeryville, CA \\
  \texttt{sadia.afroz@avast.com} \\
   \And
  Hany Farid\\
  School of Information \& EECS\\
  University of California, Berkeley \\
  Berkeley, CA \\
  \texttt{hfarid@berkeley.edu} \\
}

\begin{document}
\maketitle

\begin{abstract}
The internet promised to democratize access to knowledge and make the world more open and understanding. The reality of today's internet, however, is far from this ideal. Misinformation, lies, and conspiracies dominate many social media platforms. This toxic online world has had real-world implications ranging from genocide to, election interference, and threats to global public health. A frustrated public and impatient government regulators are calling for a more vigorous response to mis- and disinformation campaigns designed to sow civil unrest and inspire violence against individuals, societies, and democracies. We describe a large-scale, domain-level analysis that reveals seemingly coordinated efforts between multiple domains to spread and amplify misinformation. We also describe how the hyperlinks shared by certain Twitter users can be used to surface problematic domains. These analyses can be used by search engines and social media recommendation algorithms to systematically discover and demote misinformation peddlers.

\end{abstract}


\section{Introduction}

By mid-May of 2020, in the midst of the global pandemic, 28\% of Americans believed Bill Gates plans to use COVID-19 to implement a mandatory vaccine program with tracking microchips~\cite{yougov20}. Belief in this conspiracy is not unique to Americans. In global surveys~\cite{nightingale21} across Central and South America, the Middle East, Northern Africa, the United States, and Western Europe, 20\% of the public believes this bizarre claim. This conspiracy spread primarily through social media posts.

The far-reaching, far-right QAnon conspiracy alleges a cabal of Satan-worshipping cannibalistic pedophiles is running a global child sex-trafficking ring that was plotting against Donald Trump. A recent poll finds 37\% of Americans are unsure whether QAnon is true or false, and 17\% believe it to be true~\cite{ipsos20}. The conspiracy was, again, created and spread through social media posts, along with an assist from Trump himself~\cite{npr20}.

The common thread in Bill Gates' COVID-microchips, QAnon's Satan-worshipping cannibals, and the long litany of conspiracies and lies polluting the internet, is the recommendation algorithms that aggressively promote the Internet's flotsam and jetsam onto our news feeds and watch lists, plunging us into increasingly isolated echo chambers devoid of reality. 

Tackling misinformation at any scale requires striking a balance between public safety and creating an environment that allows for an open exchange of ideas. We don't necessarily advocate a specific solution to achieve this balance, but rather seek to provide the tools to help others find this balance.
 
By way of nomenclature, we will refer to a broad category of domains that traffic in conspiracies, distortions, lies, misinformation, and disinformation -- whether they are maintained by a state-sponsored actor, a private or public entity, or an individual -- as "misinformational domains." All other domains will be referred to as "informational domains." We describe in more detail, in Section~\ref{sec:dataset}, how domains are characterized as either informational or misinformational.

Tackling misinformation on a per-post/image/video basis (e.g.,~\cite{li2018exposing,matern2019exploiting,li2020celeb,agarwal2019protecting,guera2018deepfake,shu2019role,shu2019beyond,ruchansky2017csi,tschiatschek2018fake,liu2018early,hanselowski2018retrospective,zhang2019detecting,girgis2018deep,shu2018understanding,reis2019supervised,aphiwongsophon2018detecting,karimi2018multi,shu2019detecting,aldwairi2018detecting,ma-etal-2017-detect,hounsel2020identifying,tang2021down}) is leading to a maddeningly massive game of online whack-a-mole. At the same time, social networks are under intense pressure from the public and government regulators to address the scourge of misinformation. We propose that search engines and social media recommendation algorithms would benefit from more aggressively demoting entire domains that are known to traffic in lies, conspiracies, and misinformation. To this end, we describe two techniques for rooting out domains that consistently or primarily traffic in misinformation. This type of domain-level analysis might also be helpful to fact checkers evaluating the reliability of source material. 

To better understand the online misinformation ecosystem, we build two networks of misinformational and informational domains: a domain-level hyperlink network and a social-media level link sharing network. The domain-level network represents the hyperlinking relationship between domains. The social-media network represents the link-sharing behavior of social network users. From these networks, we test two main hypotheses: (1) misinformational domains are more connected (through hyperlinks) to each other than to informational domains; and (2) certain social media users are super-spreaders of misinformation. Our primary contributions include:
\begin{enumerate}
    \item Collating and curating a large set of more than 1000 domains identified as trafficking in misinformation.
    \item Revealing a distinct difference between how misinformational and informational domains link to external domains.
    \item Showing how hyperlink differences can predict if a domain traffics in misinformation.
    \item Revealing that certain Twitter users have predictable patterns in their spread of misinformation.
    \item Building a classifier for predicting the likelihood a domain is a misinformation peddler based on how specific Twitter users engage with a domain.
\end{enumerate}

\subsection{Related Work}
\label{sec:relatedwork}

Over the past five years, academic research on assessing and mitigating misinformation has increased significantly, as has the public's and government's interest in this pressing issue. Research has focused on understanding the nature of misinformation and its impact on the general population~\cite{nightingale21,lewandowsky2012misinformation,thorson2016belief}, understanding how misinformation spreads~\cite{shao2018anatomy,tang2021down,vosoughi2018spread,ma-etal-2017-detect}, and the automatic detection of misinformation~\cite{shu2019role,shu2019beyond,ruchansky2017csi,tschiatschek2018fake,liu2018early,hanselowski2018retrospective,zhang2019detecting,girgis2018deep,shu2018understanding,reis2019supervised,aphiwongsophon2018detecting,karimi2018multi,shu2019detecting,aldwairi2018detecting,hounsel2020identifying,tang2021down,KHAN2021100032}. 

With 53\% of Americans getting at least some of their news from social media~\cite{pew}, significant efforts have focused on the promotion and spread of misinformation on social media. Vosoughi et al.~\cite{vosoughi2018spread}, for example, analyzed the spread of misinformation on Twitter and found that false news spreads faster than true news. This study also found false news is more novel than true news and is designed to inspire a strong response of fear, disgust, and surprise, and hence more engagement in terms of likes, share, and retweets. Faddoul et al.~\cite{faddoul20} and Tang et al.~\cite{tang2021down} showed how YouTube's own recommendation algorithms help spread conspiracies and misinformation. Automated tools, such as Hoaxy~\cite{shao2018anatomy}, reveal in real time how misinformation spreads on Twitter.

The distinctive characteristics of false news stories make it somewhat easier to automatically detect them. A wide range of machine learning approaches have been employed including both traditional classifiers (SVM, LR, Decision Tree, Naive Bayes, k-NN) and machine learning (CNN, LSTM, Bi-LSTM, C-LSTM, HAN, Conv-HAN) models to demonstrate that false news can be automatically detected with a high level of accuracy (see~\cite{KHAN2021100032} for a comparative study of detection approaches). Having a large and accurately labeled list of misinformational news, however, is difficult to obtain, which is why most studies use small datasets.

Detection, debunking and fact-checking alone, however, are unlikely to stem the flow of online misinformation. It has been shown, for example, that the effect of misinformation may persist even after false claims have been debunked~\cite{chan2017debunking,lewandowsky2017beyond}. 

We systematically study how over 1000 domains previously identified as peddlers of misinformation, are connected with one another and how this connection can be used to detect and disrupt misinformational networks. This type of hyperlink analysis has previously been examined, however not specifically in the space of misinformation. By analyzing 89 news outlets, for example, Pak et al.~\cite{pak2020intermedia} found that partisan media outlets are more likely to link to nonpartisan media, but that liberal media link to liberal and neutral outlets, whereas conservative media link more exclusively to conservative outlets. In analyzing hyperlinks between news media between 1999 to 2006, Weber et al.~\cite{weber2012newspapers} found that establishing hyperlinks with other, younger news outlets strengthens the position of that organization in the network thus boosting traffic.

In contrast to these previous works, by analyzing significantly larger networks ($>$ 1000), we demonstrate more robust patterns of hyperlinking, and specifically focus on the growing problem of misinformation and coordinated misinformation peddlers.

\begin{table}[t]
\begin{center}
  \begin{tabular}{r|l||l|l}
    \hline
     & misinfo domain & info domain & info category \\
    \hline
    \hline   
    1  & hollywoodlife.com & espn.com & sports  \\
    2  & radaronline.com & adobe.com & business\\
    3  & newidea.com.au & fandom.com & entertainment \\
    4  & yournewswire.com & myshopify.com & business \\
    5  & madworldnews.com & bbc.com & newsandmedia\\
    6  & thecommonsenseshow.com & salesforce.com & business\\
    7  & disclose.tv & nytimes.com & newsandmedia\\
    8  & pakalertpress.com & force.com & business\\
    9  & collective-evolution.com & dailymotion.com & entertainment\\
    10  &fellowshipoftheminds.com  & primevideo.com & entertainment\\
    11  & govtslaves.info  & academia.edu & education\\
    12  & investmentwatchblog.com & yelp.com & business\\
    13  & jewsnews.co.il  & sciencedirect.com & education\\
    14  & occupydemocrats.com & mailchimp.com & business\\
    15  & twitchy.com & line.me & business\\
    16  & worldtruth.tv & deviantart.com & entertainment \\
    17  & abovetopsecret.com & quizlet.com & education \\
    18  & activistpost.com & nfl.com &  sports\\
    19  & amren.com & okta.com & business\\
    20  & amtvmedia.com & weather.com & entertainment\\
    \hline  
  \end{tabular}
\end{center}
\caption{Top-ranked misinformation and information domains in our data set.}
\label{tab:top20-sites}
\end{table}
%
%

\section{Methods}

We begin by collating and curating several public databases of previously identified misinformational and information domains. The domain-level hyperlink network is constructed by scraping all hyperlink tags ({\tt <a href="..." </a>}) from these domains. These hyperlinks can be to either an internal or external page. A level-1 scraping collects all hyperlinks from the top-level domain; a level-2 scraping collects all hyperlinks by following the level-1 links and repeating the scraping. A graph, $G=(V,E)$ is constructed from the scraped domains. Each vertex/node $v \in V$ corresponds to a domain, and each directed edge $e=(A,B) \in E$ corresponds to a hyperlink from domain $A$ to domain $B$. As described below, this graph is used to evaluate our underlying hypothesis and to gain further insight in coordinated efforts by seemingly unconnected domains.

The social-media level link sharing network is constructed using the Twitter API to find users who shared links to misinformational domains. We use a user-sharing feature vector as input to linear classifier to predict a domain as being a likely source of misinformation.

\subsection{Data Set}
\label{sec:dataset}

We begin by describing the collation and curation of four publicly available misinformation datasets.  
\begin{itemize}
\item BS Detector~\footnote{https://www.kaggle.com/mrisdal/fake-news}: This data set is based on the "BS detector" browser extension~\footnote{https://github.com/Bastlynn/bs-detector}. This extension uses a manually curated list of misinformational domains to label linked articles as reliable or not. This data set consists of 244 unique domains.
\item Columbia Journalism Review~\footnote{https://www.cjr.org/fake-beta\#methodology}: This dataset consists of manually curated misinformational stories scraped from Factcheck~\footnote{https://www.factcheck.org/2017/07/websites-post-fake-satirical-stories/}, Fake News Codex~\footnote{http://www.fakenewscodex.com/}, OpenSources~\footnote{http://www.opensources.co/}, PolitiFact~\footnote{https://www.politifact.com/} and Snopes~\footnote{https://www.snopes.com/}. This data set consists of 155 unique domains.
\item FakeNewsNet~\footnote{https://github.com/KaiDMML/FakeNewsNet}: This data set consists of manually curated misinformational stories scraped from PolitiFact~\footnote{https://www.politifact.com/} and GossipCop~\footnote{https://www.gossipcop.com/}. While PolitiFact is primarily focused on political news, GossipCop is primarily focused on the entertainment industry. This data set consists of 898 unique domains.
\item Media Bias Fact Check~\footnote{https://mediabiasfactcheck.com/fake-news/}: This data set consists of a manually curated and continually updated list of news media domains with attributes such as Factual Accuracy, Political Bias, Funding/Ownership, Country, etc. Additionally, this data set contains an evolving list of 100 websites categorized as satire, and 310 websites categorized as conspiracy-pseudoscience. This data set consists of 410 unique domains.
\end{itemize}

In total, these four data sets consist of 1,707 domains. There is, however, overlap between these datasets, which once removed yields 1,389 distinct misinformational domains. There are several limitations to immediately using these domains in our analyses. The GossipCop, FakeNewsNet, and PolitiFacts entries, for example, only provide the headline of the offending news article, from which we had to perform a reverse heuristic Google search to identify the source domain. This reverse search does not always identify the offending domain; entertainment stories, for example, often lead to domains like imdb.com and people.com.

To contend with these limitations, we applied a ranking of the 1,389 domains to down-rank mislabelled domains like imdb.com. Each domain $i$ in our original data set was assigned a score of $s_i = f_i\exp(r_i/5000)$, where $f_i$ is the frequency with which domain $i$ appeared in our original data set, and $r_i$ is the domain's Alexa top-million ranking. The exponential term weights the observation frequency so that highly ranked Alexa domains will have a nearly unit-value weight, and lower rated domains will have a higher weight. The domains with the largest scores $s_i$ were then categorized as misinformational. Despite this ranking system, a dozen clearly non-misinformational domains remained in our data set, like theonion.com and huffingtonpost.com. These domains were manually removed, yielding a total of 1,059 domains.

We paired these 1,059 misinformational domains with 1,059 informational domains corresponding to the top-ranked Alexa domains (which we manually verified are trustworthy domains). We selected 222 domains from the "news \& media" Alexa categorizaton, 198 domains from each of the "business", "education", "entertainment", and "sports" categories, "45" from "health", and 15 from "religion." Shown in Table~\ref{tab:top20-sites} are the $20$ top-ranked misinformational domains and informational domains and corresponding categories.
 
\subsection{Domain Scraping}
 
We next used OpenWPM~\footnote{https://github.com/mozilla/OpenWPM} to scrape the hyperlink on each of the 1,059 domains. The hyperlinks tag ({\tt <a href="..." </a>}) is used to link to an internal or external page. OpenWPM is used to scrape the top-level domain for each hyperlink (level 1), and to scrape all pages linked from this top-level (level 2). This scraping was performed from a Google Cloud Machine with no user login and running Ubuntu 20.04 with 4vCPUs, 16GB RAM and 500GB disk space. Before a hyperlink was scraped the browser was reset and all the cookies were deleted. This entire process was repeated once every two weeks over a six-week period between Feb 19, 2021 and Apr 2, 2021. Any domain that returned a $404$ error were excluded from our analysis, yielding a total of 874/1,059 misinformational domains and 888/1,059 informational domains.

\begin{figure}[t]
    \includegraphics[width=\textwidth]{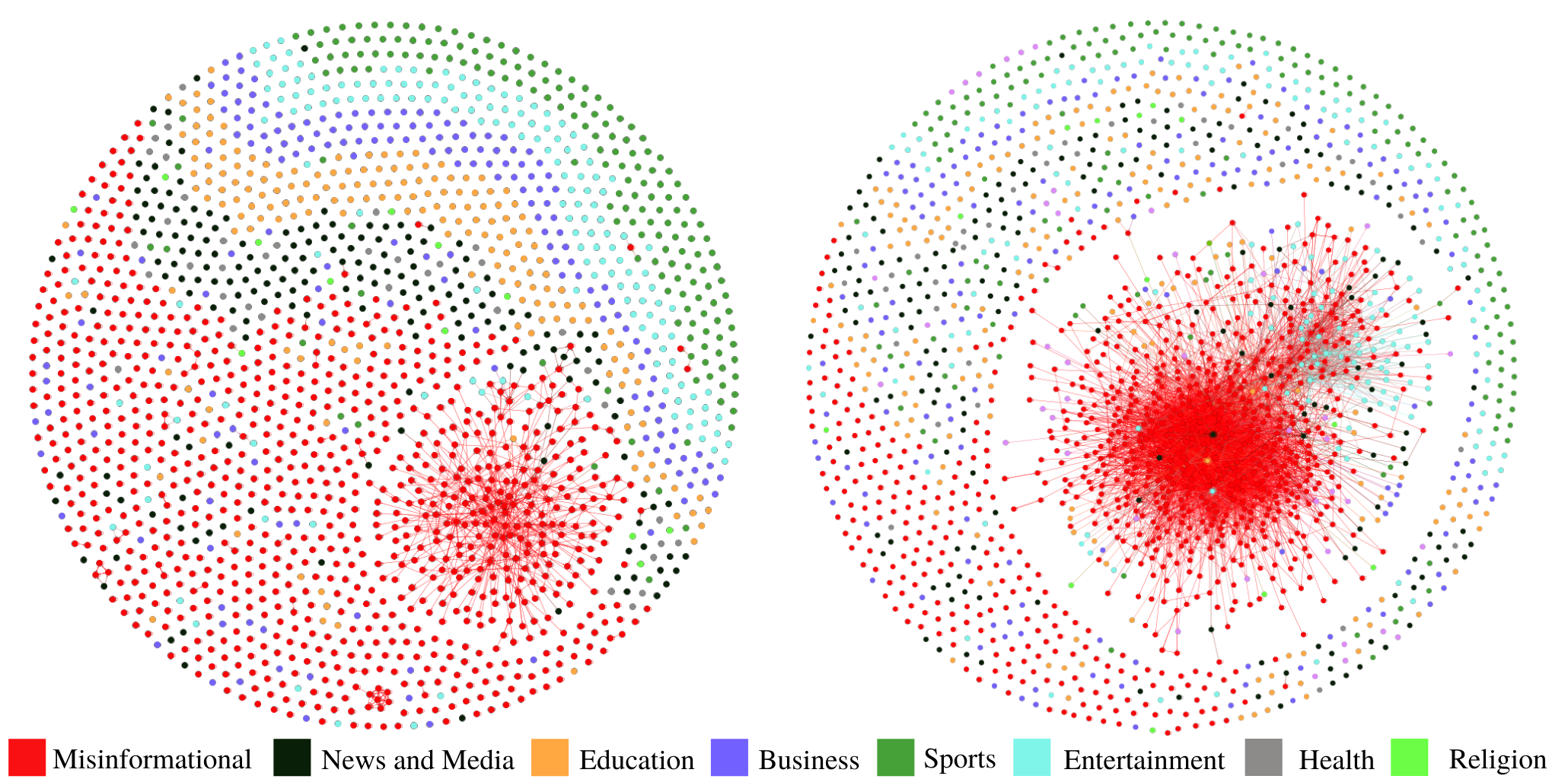}
    \caption{Hyperlink connectivity between misinformational (red) and informational domains for level-1 (left) and level-2 (right) scraping.. Each node corresponds to a domain and a directed edge between domain $i$ and domain $j$ signifies a hyperlink on domain $i$ to domain $j$. The nodes colored red correspond to misinformational domains, and the remaining nodes correspond to different categories of information domains.}
    \label{fig:graph-hyperlinks}
\end{figure}

After scraping all informational and misinformational domains, we constructed an unweighted, directed graph of hyperlinks in which the graph nodes are the domains and a directed edge connects one domain that hyperlinked to another. For example, {\tt hoggwatch.com} hyperlinks to {\tt www.infowars.com/posts/<...>} is processed as a directed edge from {\tt hoggwatch.com} to {\tt infowars.com}. Shown in Fig.~\ref{fig:graph-hyperlinks} are the level-1 (left) and level-2 (right) graphs in which we can clearly see the strong misinfo-misinfo connections and weak misinfo-info connections.
 
To visualize and analyze the large hyperlink network, we use the open-source tool Gephi~\cite{bastian2009gephi}. We use the Louvain method~\cite{newman2006modularity,blondel2008fast} for detecting communities of domains that are connected via hyperlinking relationship. This method is a fast heuristic based on modularity optimization. Networks with high modularity have dense connections between the nodes within modules but sparse connections between nodes in different modules. Since the misinformational domains are more connected to each other than informational domains, we hypothesize the misinformational and informational domains will belong to different communities.   

%
%

\section{Results: Domain-Level Hyperlinking}
\label{sec:analysis}

\subsection{Overview}

Each domain in the network created from the misinformational and informational domains are assigned a label of "misinfo", "info" or "none". The "none" categorization is used to classify domains not in the misinformational or informational data set. Shown in the top portion of Table~\ref{tab:hyperlinks} is the number and proportion (\%) of level-1 hyperlinks from misinfo and info (rows 1-2) to misinfo, none, and info (columns 1-3). Here we see a huge difference, with 17.90\% of hyperlinks of misinfo domains linking to misinfo domains, as compared to only 0.62\% for info domains. Similarly, albeit a smaller effect, 4.37\% of hyperlinks on misinfo domains are to info domains, as compared to 13.45\% for info domains.

Shown in the lower portion of Table~\ref{tab:hyperlinks} is the distribution of hyperlinks for level-2 hyperlinks. A similar pattern emerges, albeit not quite as dramatic: 9.27\% of hyperlinks on misinfo domains link to misinfo domains, as compared to only 1.03\% for info domains, and 7.31\% of hyperlinks on misinfo domains are to info domains, as compared to 9.78\% for info domains. Also shown in this table is the breakdown of links based on domain categories. Here we see that in level-1 and level-2 "entertainment", "news \& media", and "religion" are more likely to link to misinfo domains.

Included in the 60 info to misinfo hyperlinks we discovered, is the news \& media site {\tt drudgereport.com} hyperlinking to sites like {\tt breitbart.com} and {\tt thegatewaypundit.com}, each of whom have been implicated in spreading misinformation and conspiracies~\cite{poltifact2020jul,leadstories2020}. Other examples include the entertainment site {\tt indiewire.com} linking to {\tt hollywoodlife.com}, notorious for spreading gossip and misinformation.

\begin{table}[t]
\begin{center}
  \begin{tabular}{r|cc|cc|cc||c}
    \hline
    &\multicolumn{2}{c|}{misinfo} & \multicolumn{2}{c|}{none} & \multicolumn{2}{c||}{info} & \multicolumn{1}{c}{total}\\
    & \#links & \% & \#links & \% & \#links & \% & \#links \\
    \hline
    misinfo  & 1404 & 17.90 & 6098 & 77.74 & 343 & 4.37 & 7844 \\
    info     & 60 & 0.62  & 8282 & 85.94 & 1296 & 13.45 & 9637 \\
    \hline
    entertainment     & 21 & 1.47 & 1209 & 84.66 & 198 & 13.87 & 1428 \\
    education     & 4 & 0.27 & 1204 & 82.64 & 249 & 17.09 & 1457 \\
    news\&media     & 28 & 1.10 & 2211 & 87.12 & 300 & 11.82 & 2538 \\
    business     & 2 & 0.15 & 1165 & 85.22 & 201 & 14.70 & 1367 \\
    sports     & 5 & 0.20 & 2170 & 88.10 & 289 & 11.73 & 2463 \\
    religion     & 1 & 0.82 & 107 & 87.70 & 14 & 11.48 & 122 \\
    health     & 0 & 0 & 217 & 82.20 & 47 & 17.80 & 264 \\
    \hline
    \hline   
    misinfo  & 6595 & 9.27 & 56325 & 82.98 & 4961 & 7.31 & 67881 \\
    none     & 1453 & 2.59 & 51362 & 91.65 & 3228 & 5.76 & 56043 \\
    info     & 557 & 1.03 & 48008 & 90.38 & 5264 & 9.78 & 53828 \\
    \hline
    entertainment & 252 & 3.16 & 6799 & 85.28 & 923 & 11.58 & 7973\\
    education & 113 & 0.90 & 11217 & 89.00 & 1274 & 10.11 & 12604\\
    news\&media & 106 & 0.81 & 11888 & 91.31 & 1027 & 7.89 & 13020\\
    business & 32 & 0.38 & 7542 & 90.38 & 771 & 9.24 & 8345 \\
    sports & 44 & 0.43 & 9115 & 89.23 & 1056 & 10.34 & 10215\\
    religion & 7 & 1.55 & 389 & 86.06 & 57 & 12.61 & 452\\
    health & 4 & 0.33 & 1054 & 86.82 & 156 & 12.85 & 1214\\
    \hline  
  \end{tabular}
\end{center}
\caption{Level 1 (top) and level 2 (bottom) hyperlink network analysis. Each entry corresponds to the number and percentage of total links from one domain (row) to another (column). The label "none" corresponds to domains that are not labeled as misinformational (misinfo) or informational (info). In level 1 analysis, 17.90\% of hyperlinks on misinfo domains are to other misinfo domains, as compared to only 0.62\% of hyperlinks on info domains. This pattern persists in level 2, albeit with a smaller difference, 9.27\% versus 1.03\%.}
\label{tab:hyperlinks}
\end{table}

\subsection{Insights}

\begin{figure}[t]
    \includegraphics[width=\textwidth]{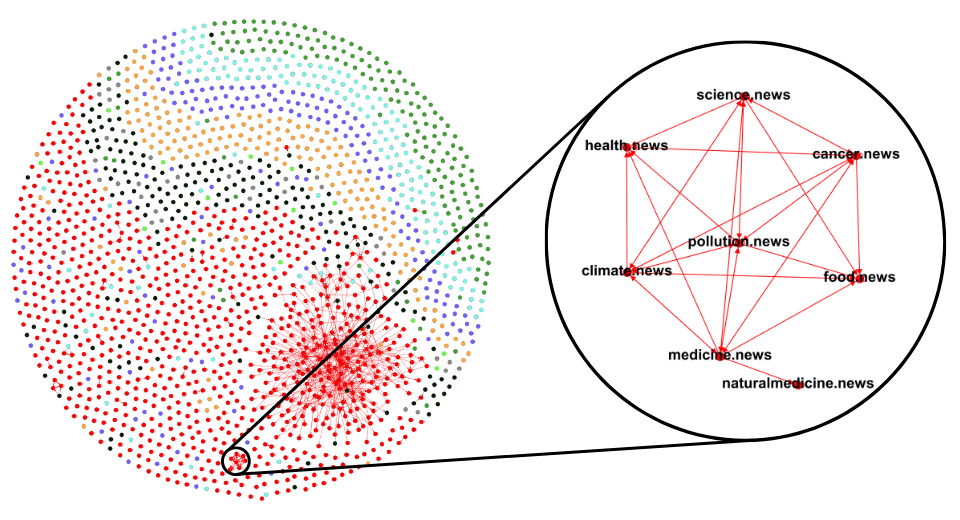} 
    \caption{A magnified view of eight almost fully connected {\tt .news} domains, all owned by Webseed.}
    \label{fig:news-mike}
\end{figure}

Shown in the hyperlink graph in Fig.~\ref{fig:news-mike} is a small, nearly fully-connected clique of eight domains: {\tt cancer.news}, {\tt climate.news}, {\tt food.news}, {\tt health.news}, {\tt medicine.news}, {\tt naturalmedicine.news}, {\tt pollution.news}, and {\tt sciences.news}. This clique is an example of how individual domains can amplify misinformation by, disproportionately, linking to like-minded domains.

Following a reverse whois lookup ({\tt www.whois.com}), we find all of these misinformational domains are owned by Webseed, LLC based in Arizona, USA. A deeper internet search reveals this LLC was created by Mike Texas, a pseudonym for Mike Adams, founder of Natural News. According to Wikipedia "Natural News (formerly NewsTarget, which is now a separate sister domain) is an anti-vaccination conspiracy theory and fake news website known for promoting pseudoscience and far-right extremism. Characterized as a ``conspiracy-minded alternative medicine website", Natural News has approximately 7 million unique visitors per month."~\cite{naturalNews}

\begin{figure}[t]
    \begin{center}
        \includegraphics[width=0.75\textwidth]{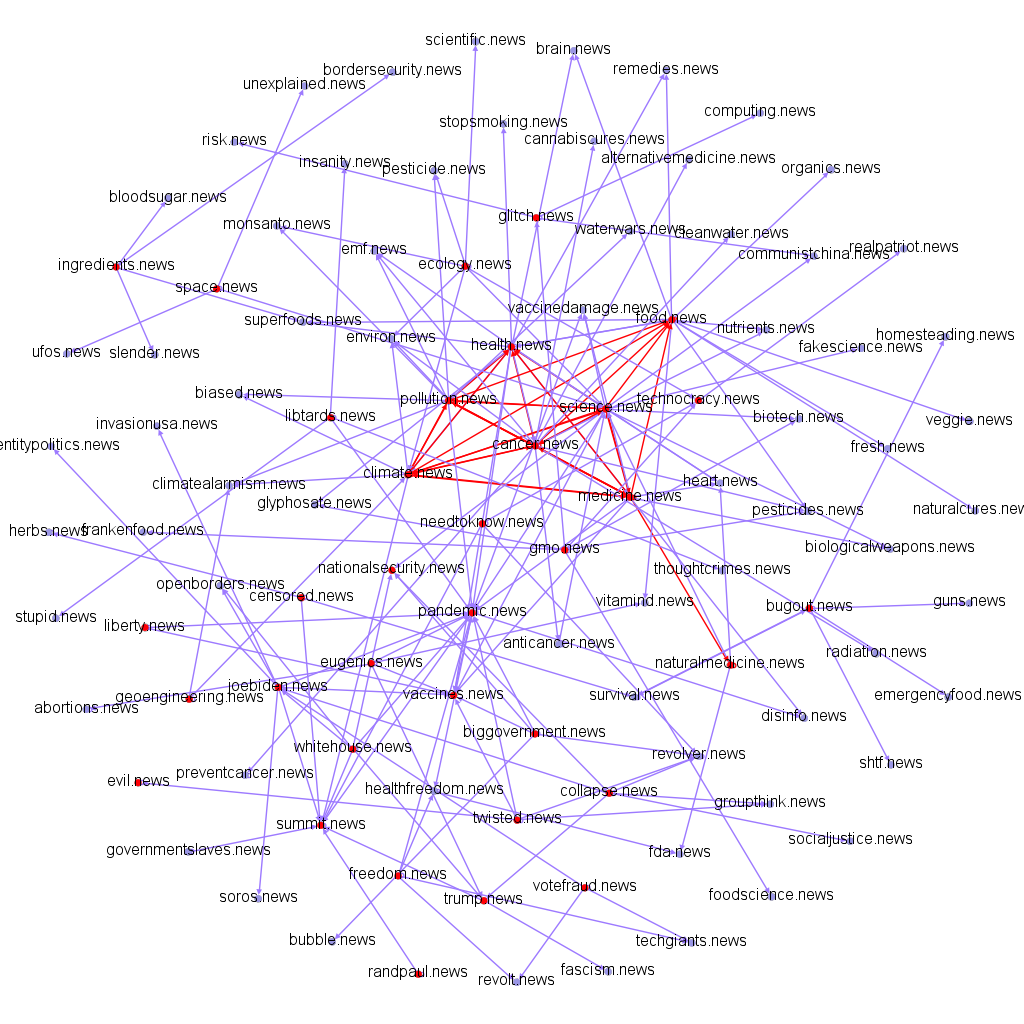} 
    \end{center}
    \caption{Level-2 scraping of hyperlinks reveals a network of $102$ {\tt .news} domains. The red nodes and edges correspond to the original eight-node clique, Fig.~\ref{fig:news-mike}. The remaining red nodes are those from our original misinformational data set, and the magenta nodes/edges are misinformational domains discovered by the second-level scraping.}
    \label{fig:news-mike-level2}
\end{figure}

The level-2 scraping reveals this clique of eight domains is the tip of the iceberg. Shown in Fig.~\ref{fig:news-mike-level2} is a large network of $102$ {\tt .news} domains all owned by Webseed, LLC. In this figure, the red nodes and edges correspond to the original eight-node clique, Fig.~\ref{fig:news-mike}. The remaining red nodes are those from our original misinformational data set, and the magenta nodes/edges are misinformational domains discovered by the level-2 scraping. This analysis shows the power of the hyperlinking theory to discover new domains peddling in misinformation.

Our level-1 scraping also discovered a smaller clique of three domains: {\tt blackeyepolitics.com}, {\tt greatamericandaily.com} and {\tt americanpatriotdaily.com}. Despite forming a fully-connected clique, at first glance these domains appear to be unrelated each with the following ownerships: Rising Media News Network LLC, Great American Daily Press LLC, and American Patriot News LLC, respectively. These domains, however are all owned by David A. Warrington~\cite{engadget19}. Warrington also owns other domains including {\tt conservativerevival.com} and {\tt liberalpropagandaexposed.com}, the former of which we discovered through our level-2 scraping. This analysis reveals how mutual hyperlinking can reveal seemingly coordinated misinformation efforts despite owners' efforts to conceal their coordination.

The above insights were gained by visually inspecting the hyperlink graphs in Fig.~\ref{fig:graph-hyperlinks}. As these graphs increase in size, however, this type of manual approach will quickly become impractical. We, therefore, employ a community detection algorithm~\cite{blondel2008fast} to discover connections between a subset of domains. 

We applied this community detection algorithm to the level-1 graph in Fig.~\ref{fig:graph-hyperlinks}. This analysis revealed two communities. The first consisted of the following eight domains: 
{\tt globalresearch.ca}, {\tt journal-neo.org}, {\tt sott.net}, {\tt strategic-culture.org}, {\tt swprs.org}, {\tt theduran.com}, {\tt thelibertybeacon.com}, and {\tt wikispooks.com}. The average degree -- defined as the sum of all the edges incident to a node -- of this community was 3.12 and the graph density was 0.446 (a graph density of 1 signifies a complete graph). Some of these domains have previously been identified as spreading misleading and false COVID-19 related information~\cite{healthfeedback2020,bbc2020,insider20206}, three of which appear to be controlled by the Russian intelligence agency~\cite{nytimes2020,cbc2020,thebulletin}.

The second community consisted of the following five domains: {\tt cnsnews.com}, {\tt protrumpnews.com}, {\tt thegatewaypundit.com}, {\tt thepoliticalinsider.com}, and {\tt waynedupree.com}. The average degree of this community was 2.11 and the graph density was 0.12. These domains focus primarily on pro-Trump misinformation. The site {\tt thegatewaypundit.com}, led by Jim Hoft, for example, promoted false rumors about voter fraud and Hillary Clinton's health in the 2016 US-national election. Earlier this year, Hoft was banned from Twitter for "repeated violations of Twitter's civic integrity policy."  Similarly, {\tt cnsnews.com} is run by the daughter of  Republican mega-donor, Robert Mercer. Mercer has been implicated in the weaponization of millions of misinformation-spreading Twitter bots~\cite{dailymail2017}. 

These types of communities are quite common. Our analysis revealed misinformational domains dominate the five largest communities: (1) the largest community consisted of 79 domains, 88.6\% of which are misinformational; followed by (2) 75 domains, 46.7\% of which are misinformational; (3) 50 domains, 38\% of which are misinformational; (4) 46 domains, 87.0\% of which are misinformational; and lastly (5) 39 domains, 82.1\% of which are misinformational.

\section{Results: Link Sharing on Social Media}

Domain cross-linking among misinformational domains adds to the spread of lies and conspiracies. Additionally, billions of world-wide, social-media users are at least equally responsible for spreading misinformation on social media. A recent report, for example, Facebook's own internal research found 111 users are responsible for the majority of anti-vaccination misinformation~\cite{dwoskin21}.

We investigate the ability to identify misinformational domains by tracking the hyperlinks shared by certain social-media users. Because of the relative ease of access, we focus on Twitter's publicly available user data. In particular, we enlist two Twitter APIs: (1) The Search Tweets API\footnote{https://developer.twitter.com/en/docs/twitter-api/v1/tweets/search/guides/standard-operators} allows filtering tweets based on a query term against a tweet's keywords, hashtags, or shared URLs. We filter tweets by matching shared URLs against our misinfo/info URL dataset, surfacing which users are sharing a particular domain.; and (2) The Get Tweet Timelines API\footnote{https://developer.twitter.com/en/docs/twitter-api/v1/tweets/timelines/api-reference/get-statuses-user\_timeline} allows querying all tweets surfaced from the Search Tweets API. In our case, we extract the domain URLs shared by the Twitter users surfaced in the previous step. Although we don't consider them here, the data returned by both APIs contains geo-location, replied-to, time, and other attributes that could be leveraged in the future.

\begin{figure}[t]
   \begin{center} 
      \includegraphics[width=0.75\textwidth]{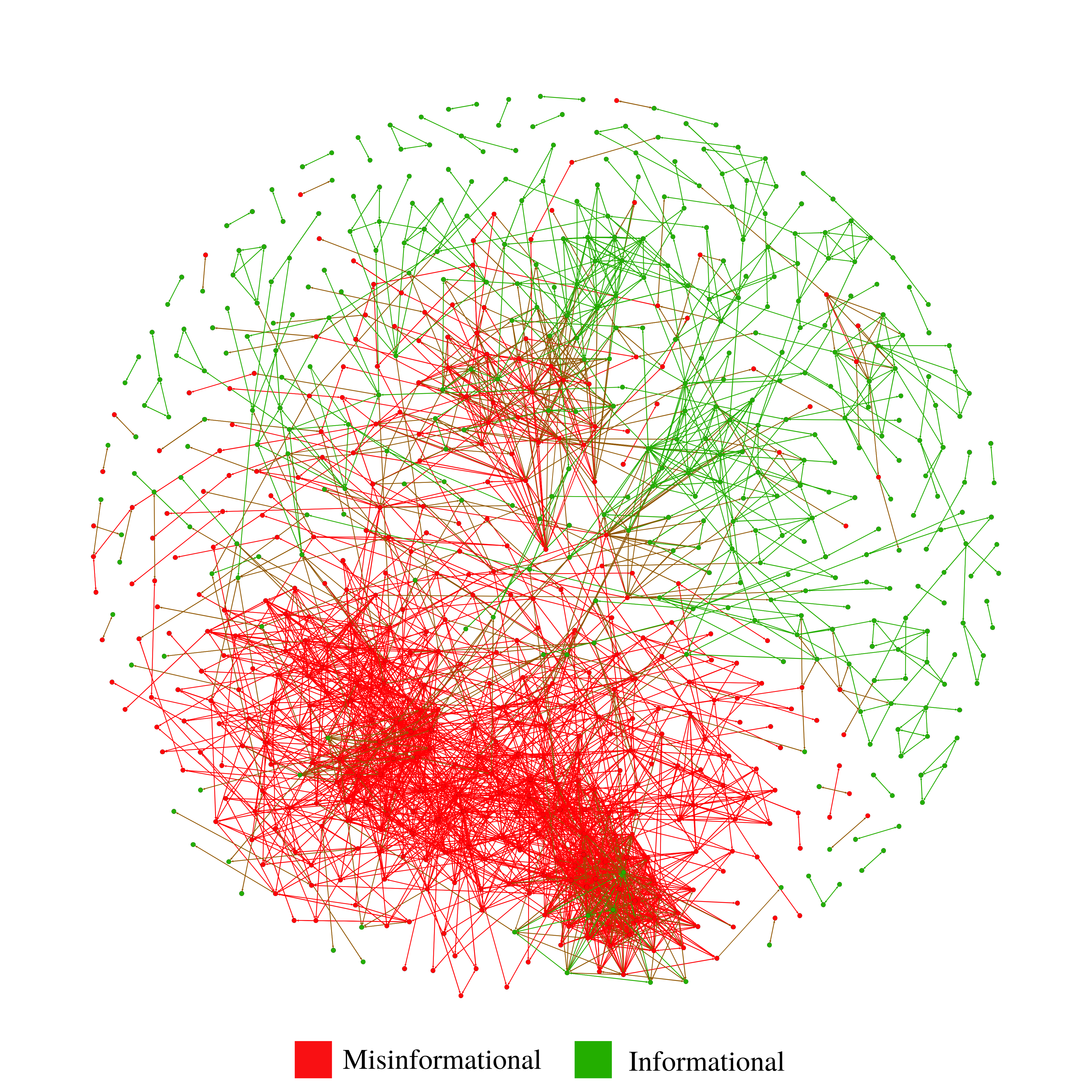} 
    \end{center}
    \caption{Mutual domain URL sharing by Twitter users, where each node is a domain. Undirected edges connect two domains $A$ and $B$ if at least 1\% of the users sharing either $A$ or $B$ also shared the other domain (i.e.,~the Jaccard index is greater than 1\%). Nodes with no connections have been dropped.
}
    \label{fig:link-sharing-plot}
\end{figure}

Each domain in our misinformational and informational data set is represented by a binary-valued vector corresponding to whether a particular user shared the domain URL. In order to avoid an overly large and sparse representation, starting with a total of 289,984 users from our initial Twitter search, we eliminated 244,525 users who shared less than 2 domains each. The final 45459-D, binary-valued vector serves as the feature vector for each domain. 

We further remove any domains with fewer than $5$ total tweeters, yielding a reduction from 961 to 451 misinformational domains and 962 to 705 informational domains. A total of 1156 domains are split into a $75\%/25\%$ training/testing split. In order to balance the training data, random oversampling with replacement is applied to the minority (misinformational) class.

We trained a logistic regression (LR) using 75\% of the data and then evaluated it on the rest of the 25\% data set. The LR hyperparameters are tuned to maximize the F1-score. On testing, the LR classifier, with a support of 113/176 misinfo/info domains, achieved an F1-score of $0.75/0.85$, with precision $0.78/0.84$ and recall $0.73/0.87$. Domains such as {\tt dailywire.com}, {\tt fortherightnews.com}, {\tt newsblaze.com}, {\tt nickiswift.com}, {\tt therightscoop.com}, {\tt trueactivist.com}, and {\tt usasupreme.com}, were correctly classified as misinformational. The classifier incorrectly inferred the following domains as informational, indicating they are shared along with other informational domains: {\tt cancer.news}, {\tt celebrityinsider.org}, {\tt hollywoodreporter.com}, {\tt medicine.news},  {\tt medicalmedium.com}, {\tt science.news}, and {\tt trump.news}.

To visualize the informational and misinformational domains shared by Twitter users, we construct an undirected graph $G = (V, E)$ where each node $v \in V$ represents a domain, and two vertices $A, B \in V$ are connected by an undirected edge, $e = (A, B) \in E$, if at least 1\% of the users sharing either domain $A$ or $B$ share both of them (i.e.,~Jaccard Similarity $\ge$ 1\%). In this visualization, Fig.~\ref{fig:link-sharing-plot} -- and consistent with our findings from the previous section -- we see strong misinfo-misinfo connectivity and weak misinfo-info connectivity. In particular, misinfo/info domains have an average of 7.55/1.57 connections, respectively, within their category, and 1.00/0.86 connections outside their category.

A community analysis reveals some large communities dominated by misinformation domains: (1) the largest such community consists of 131 domains, 89.3\% of which are misinformational; followed by (2) 58 domains, 86.2\% of which are misinformational; and (3) 46 domains, 91.3\% of which are misinformational. We found a particularly interesting community of 14 domains dominated by climate-change deniers, which includes domains like:  {\tt cfact.org}, {\tt climatism.blog}, {\tt climatechangedispatch.com}, {\tt iceagenow.info}, {\tt iowaclimate.org}, {\tt notrickszone.com}, {\tt realclimatescience.com}, and {\tt wattsupwiththat.com}. A total of 13 of the 14 domains in this community are misinformational domains, and one in particular, {\tt wattsupwiththat.com}, has been identified by climatologist Michael E. Mann as "the leading climate change denial blog"~\cite{mann2013hockey}.


\section{Discussion}
\label{sec:discussion}

From the nature of lies, conspiracies, and rumors, to the methods for their delivery and spread, misinformation has, and is likely to continue to be an ever-evolving phenomena. While misinformation is not new, the consequences of its collision with a vast digital landscape has led to significant offline harms to individuals, marginalized groups, societies, and our very democracies. Addressing these harms will require a multi-faceted approach from thoughtful government regulation, to corporate responsibility, technological advances, and education.

As with most aspects of cybersecurity, technological solutions to addressing misinformation will themselves have to be multi-faceted. With some 500 hours of video uploaded to YouTube every minute, and over a billion posts to Facebook each day, the massive scale of social media makes tackling misinformation an enormous challenge. We propose that in conjunction with complementary approaches to tackling misinformation, addressing misinformation at the domain level holds promise to disrupt large-scale misinformation campaigns. Previous studies have found a relatively small group of individuals are responsible for a disproportionate number of lies and conspiracies. Identifying this group, and reducing their reach -- while not necessarily silencing them entirely -- holds the potential to make a large dent in the online proliferation of harmful misinformation.

We understand and appreciate the need to balance an open and free internet, where ideas can be debated, with the need to protect individuals, societies, and democracies. Social media, however, cannot hide behind the facade they are creating a neutral marketplace of ideas where good and bad ideas compete equally. They do not. It is well established that social media's recommendation algorithms favor the outrageous and conspiratorial because it increases engagement and profit. As a result, Brandies' concept that the best remedy for falsehoods is more speech, not less, simply doesn't apply in the era of algorithmic curation and amplification. We propose that identified misinformation peddlers not necessarily be banned or de-platformed, but that their content simply be demoted in favor of more honest, civil, and trustworthy content.

As with any inherently adversarial relationship, all approaches to addressing misinformation -- including ours -- will have to adapt to new and emerging threats. In our case, misinformation peddlers may add decoy hyperlinks to external trustworthy domains to escape being classified based on their hyperlinks to other misinformational domains. This, in turn, will require techniques to root out such decoy links. And so on, and on, and on. While such a cat and mouse game can be frustrating, the end game will be that it will become increasingly more difficult and time consuming to create and spread misinformation, with the eventual goal of discouraging most, leaving us to contend with the die-hard adversary. While this is not a complete success, it will mitigate the risk of misinformation and, hopefully, return some civility and trust to our online ecosystems.

\section*{Acknowledgments}

This work was supported by funding from funding Avast, Inc. (Sehgal and Farid). We thank Juyong Do and Rajarshi Gupta for their thoughtful comments and discussions.

\bibliographystyle{hacm}
\bibliography{main}


\end{document}